# The genetic and developmental enigma of rhizomes: crucial traits with limited understanding

Hongfei Chen[1*], Jenn M. Coughlan[1*]

[1]Department of Ecology and Evolutionary Biology, Yale University, New Haven, CT 06520 USA

[*]Corresponding Author: hongfei.chen@yale.edu; jennifer.coughlan@yale.edu

## *Abstract*

Rhizomes, horizontal underground stems, play fundamental roles in plant evolution, persistence, and environmental adaptation by enabling clonal propagation, resource storage, and stress resilience. Despite their ecological and agronomic importance across diverse plant lineages, the genetic and developmental regulation of rhizomes remains poorly characterized. Here, we synthesize findings from *in vitro* induction studies, *in vivo* physiological and developmental analyses, quantitative trait loci (QTL) mapping, comparative transcriptomics, and limited functional studies to evaluate current knowledge and highlight outstanding questions in rhizome biology. Results from both *in vitro* and whole-plant studies show that phytohormones, particularly auxin, cytokinin, and gibberellin, are central regulators of rhizome initiation and growth, with effects mediated in a context-dependent manner through interactions with environmental and developmental cues. Across rhizomatous species, traits such as rhizome initiation, branching, and elongation are typically under polygenic control, although comparatively simpler genetic architectures have been documented in emerging model systems like *Mimulus*. Transcriptomic analyses further highlight hormone signaling, stress-response, and carbohydrate metabolism pathways as key regulatory components. However, few genes have been functionally validated, underscoring the need for experimentally tractable systems for genetic dissection. Perennial *Mimulus* species are proposed as promising models for rhizome research due to their experimental accessibility, ecological relevance, and established genomic resources. Integrated approaches leveraging fine-mapping, near-isogenic lines, multi-omics network reconstruction, and genome editing are poised to accelerate the discovery of causal loci and regulatory networks underlying rhizome development, thereby illuminating key processes involved in plant adaptation, perenniality, and invasiveness, with direct implications for evolutionary biology and crop improvement.

# *Introduction*

A rhizome is a type of modified stem that develops from axillary buds (Gizmawy et al., 1985) and grows horizontally underground (Fig. 1a) (Guo et al. 2021). A stolon is similar to a rhizome but grows above ground (Finch et al. 2014). According to a study in *Sorghum halepense*, axillary buds on seedling shoots undergo two orientation reversals, resulting in a positional gradient along the shoot, in which the lowermost buds oriented downward developed into rhizomes, whereas the more acropetally oriented buds developed into tillers (Gizmawy et al. 1985). Similar bud reorientation and rhizome initiation were also observed in *Agropyron repens* (McIntyre 1967). The bud reversals were found to be driven by variations in radial growth rates between the internodes immediately above and below the bud, as well as by different cell division rates between the abaxial and adaxial sides of the bud (Gizmawy et al. 1985). Axillary buds on rhizomes can further either form new rhizomes or aerial shoots (Fisher 1972; Yoshida et al. 2016).

Rhizomes hold substantial biological, ecological, and economic importance (Fig. 1b). Biologically, rhizome propagation is a common mode of asexual reproduction for perennial plants due to the rhizome's capacity to generate new rhizomes, shoots, and roots (Dong and Pierdominici 1995; Duncan 1935; Huang et al. 2022). Many plants, such as bamboo (Kotangale et al. 2025; Uchimura 1980) and sheepgrass (Yang et al. 1995), rely primarily on rhizomes for natural propagation. On the other hand, the propagation through rhizomes can contribute to the weedy nature of certain plant species, such as Japanese knotweed and mugwort (Weston et al. 2005). Rhizomes are also a defining feature of perennial plants, functioning as storage organs for nutrients and energy that sustain them through winter dormancy (Chung and Kim 2012; Dohleman et al. 2012; Mitros et al. 2020). This physiological reserve underpins whole-plant persistence of perennial herbs under a broad range of environmental stresses, including drought and cold, and enables regrowth following stress release (Guo et al. 2021; Ma et al. 2020; Tejera-Nieves and Walker 2023). Consequently, compared to annual crops, perennial counterparts have some competitive advantages, including longer growing seasons and deeper root systems that tap water and nutrients at greater depths (Fan et al. 2020). Therefore, the identification of genes responsible for rhizome initiation and development is of utmost importance for crop breeding purposes, as rhizomes are integral to perenniality (Fan et al. 2020). From an ecological perspective, rhizomes play a crucial role in preventing soil erosion by effectively binding soil particles with plant roots, as supported by various studies (Balasuriya et al. 2018; Gyssels et al. 2005; Xue et al. 2016). This role is particularly important in the context of increasing global soil degradation. Rhizomes also hold economic significance, as they possess edible and medicinal properties, particularly in certain plant species. For instance, the rhizomes of ginger, turmeric and lotus are frequently utilized in various culinary applications, owing to their unique flavors and nutritional benefits. Furthermore, the rhizomes of medicinal plants such as S*alvia miltiorrhiza* (known as Danshen in Chinese) (Chong et al. 2019), *Panax ginseng* (Zhang et al. 2023), and *Atractylodes macrocephala* Koidz (known as Baizhu in Chinese) (Yang et al. 2021) are widely used in the treatment of various diseases, especially in Eastern Asian countries.

Despite the vital role of rhizomes in multiple areas, the regulation, molecular basis, and evolution of rhizomes are still poorly understood. This knowledge gap largely stems from the absence of stolons or rhizomes in well-established model plant systems, such as *Arabidopsis* and tomato (Guo et al. 2021). This review aims to synthesize current research on rhizome development, underlying molecular mechanisms, and their evolutionary significance, and to highlight promising directions for future studies in rhizome biology.

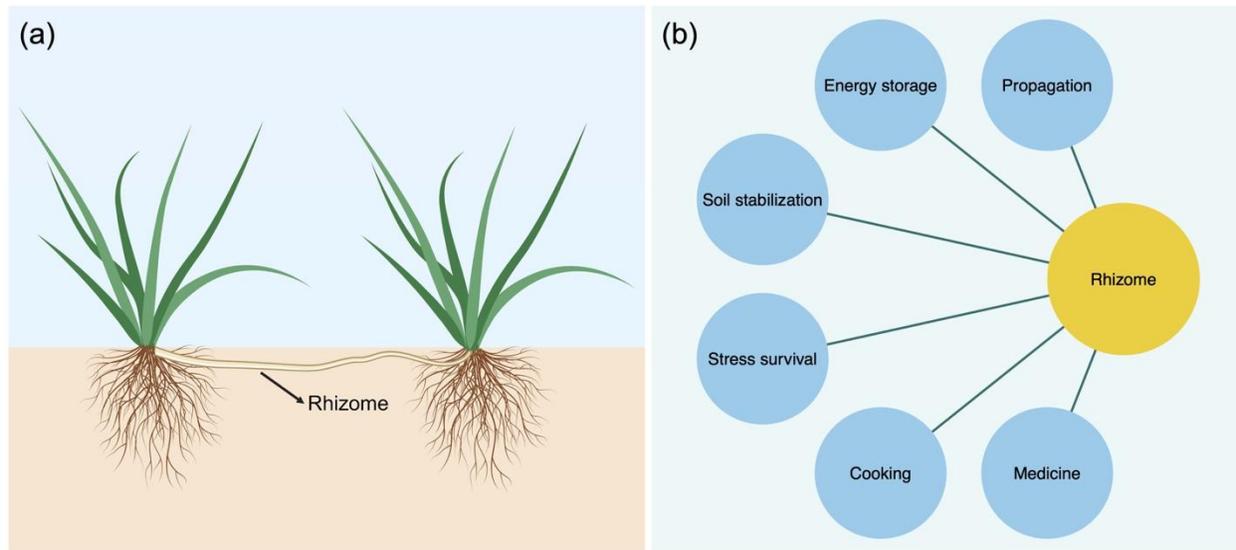

**Fig.1 Structural and functional significance of rhizomes in plants.** (a) Illustration of clonal propagation via rhizomes in a typical grass species. (b) Key ecological and anthropogenic functions of rhizomes, including roles in propagation, energy storage, stress survival and human use.

## *Regulation factors of rhizome formation and growth*

Rhizome formation and growth are shaped jointly by phytohormonal regulation and environmental cues. *In vitro* induction systems have clarified many of the hormonal requirements underlying rhizome initiation and growth, whereas *in vivo* studies reveal how these hormonal effects operate within whole-plant physiological and environmental contexts. Below, we summarize key findings from both approaches to provide an integrated view of rhizome regulation (Table 1a and 1b).

## *Insights from in vitro studies*

*Regulation factors of rhizome formation*

Cytokinin (CK) and auxin (Aux) are central regulators of rhizome initiation. CK alone can trigger rhizome formation in *Cyperus serotinus* (Omokawa et al. 1992) and in several Zingiberaceae species, including *Curcuma aromatica* (Nayak 2000) and *Zingiber officinale* (Abbas et al. 2014), whereas in orchids, Aux alone is sufficient and CK exerts little influence (Paek and Yeung 1991; Shimasaki and Uemoto 1991). In most systems, a finely balanced CK–Aux ratio is required for successful rhizome induction (Kapoor and Rao 2006; Roy and Banerjee 2002; Shimasaki and Uemoto 1991; Zahid et al. 2021). Additional hormones, including gibberellic acid (GA) (Escalante and Langille 1995; Kapoor and Rao 2006), abscisic acid (ABA) (Kim et al. 2022), jasmonic acid (JA) (Rayirath et al. 2011), and ethylene (ETH) (Rayirath et al. 2011; Shimasaki 1993) act either downstream of or in concert with CK and Aux to modulate this developmental switch. Environmental conditions also influence rhizome formation, with high sucrose concentrations (Fan et al. 2022; Gezahegn et al. 2024; Kapoor and Rao 2006; Labrooy et al. 2020;

Zahid et al. 2021), favorable photoperiod (Nayak 2000; Rout et al. 2001; Yoshida et al. 2016), optimal temperature (Maria De Fatima and Valio 1976), and mild drought stress (Almeida et al. 2005) promoting induction.

*Regulation factors of rhizome elongation and branching*

CK and Aux both promote rhizome elongation in *Gastrodia elata* (Hsieh et al. 2022). Gibberellic acid (GA) similarly enhances elongation in *Caulerpa prolifera* (Jacobs and Davis 1983) and *Solanum tuberosum* (Escalante and Langille 1998), whereas ethylene (ETH), although generally inhibitory (Vaseva et al. 2018), can stimulate elongation in *Rheum rhabarbarum* (Rayirath et al. 2011) under certain conditions. Higher sucrose concentrations also promote rhizome elongation in *Acorus calamus*, *Bambusa bambos* and *Oryza longistaminata* (Fan et al. 2022; Kapoor and Rao 2006; Subramani et al. 2014).

Aux typically stimulates rhizome branching in *Cymbidium aloifolium* (Nayak et al. 1998), *Cymbidium forrestii* (Paek and Yeung 1991), and *Geodorum densiflorum* (Roy and Banerjee 2002), whereas CK tends to inhibit it in orchid species such as *Cymbidium kanran* (Shimasaki 1995) and *C. forrestii* (Paek and Yeung 1991). Depending on concentration and species, CK can either counteract (Roy and Banerjee 2002) or enhance Aux-driven branching (Fukai et al. 2000). Beyond Aux and CK, ETH promotes branching in *R. rhabarbarum* (Rayirath et al. 2011) and *C. kanran* (Shimasaki 1995).

*Regulation factors of rhizome shooting and rooting*

CK strongly induces rhizome shoot formation across *Alstroemeria* (Khaleghi et al. 2008; Shahriari et al. 2012), *A. calamus* (Tikendra et al. 2022), *C. kanran* (Shimasaki 1995; Shimasaki and Uemoto 1990), *C. aloifolium* (Nayak et al. 1998), *G. densiflorum* (Roy and Banerjee 2002), and *Kaempferia galang* (Vincent et al. 1992). Aux similarly promotes shoot initiation in *Alstroemeria* (Hamidoghli et al. 2007; Khaleghi et al. 2008; Shahriari et al. 2012) and *G. densiflorum* (Roy and Banerjee 2002) but inhibits it in *A. calamus* (Tikendra et al. 2022). GA enhances rhizome shooting in *Zantedeschia* (Kozłowska et al. 2007) and *R. rhabarbarum* (Rayirath et al. 2009) yet suppresses it in *A. repens* (Rogan and Smith 1976). ETH likewise inhibits rhizome shoot development in *C. kanran* (Ogura-Tsujita and Okubo 2006; Shimasaki 1995) and *Kohleria eriantha* (Almeida et al. 2005), while promoting it in *Z. officinale* (Furutani et al. 1985). Environmental cues, including sucrose (Paek and Yeung 1991; Zahid et al. 2021), temperature (Leakey et al. 1978), and drought (Rice et al. 1996), also exert strong effects on rhizome shoot induction.

CK promotes rhizome rooting in *Valeriana jatamansi* (Nazir et al. 2022) but inhibits it in *Alstroemeria* (Gabryszewska and Hempel 1985; Kristiansen et al. 1999), *G. densiflorum* (Roy and Banerjee 2002), and *Ruppia maritima* (Koch and Durako 1991). Similarly, Aux promotes rooting in *Alstroemeria* (Hamidoghli et al. 2007; Khaleghi et al. 2008; Kristiansen et al. 1999), *Podophyllum hexandrum* (Nadeem et al. 2000), *Posidonia oceanica* (Balestri and Lardicci 2006), *Scopolia parviflora* (Kang et al. 2004), and *V. jatamansi* (Nazir et al. 2022) but suppresses it in *Ruppia maritima* (Koch and Durako 1991). GA inhibits root initiation in *Convolvulus sepium* (Wells and Riopel 1972), whereas sucrose enhances rooting in *Alstroemeria* (Gabryszewska 1996) and *Z. officinale* (Zahid et al. 2021).

*Synthesis and future directions for in vitro research*

Collectively, *in vitro* studies reveal that rhizome formation and growth are regulated primarily by an Aux–CK axis, with GA, ETH, and sucrose modulating its outcome under specific environmental contexts. However, because most experiments expose isolated tissues to exogenous treatments under simplified conditions, they remain largely descriptive and offer limited insight into how these signals are integrated within intact plants. Future in vitro work should therefore move beyond simple induction assays toward more mechanistic designs, employing more precisely parameterized hormonal and environmental manipulations together with molecular and cellular readouts. Such approaches will more effectively resolve causal regulatory relationships and generate testable hypotheses about conserved versus context-dependent regulatory modules for evaluation *in vivo*.

## Insights from in vivo studies

*Meristem fate determination and rhizome initiation*

In *O. longistaminata*, single-cell and spatial transcriptomic analyses identified meristematic initiation cells within a sunken parenchyma zone at the internode base as the starting point for rhizome initiation, and trajectory analysis further revealed that rhizomes originate de novo through cell fate reprogramming (Lian et al. 2024). Complementary morphological evidence indicates that axillary bud shape predicts developmental fate: dome-shaped buds penetrate the leaf sheath to produce rhizomes, whereas flat buds remain enclosed and develop into tillers (Wang et al. 2024b).

*Nutritional and environmental control*

Environmental and nutritional factors exert strong, species-specific influences on rhizome initiation and elongation across clonal plants. A series of studies in *A. repens* established that increased water availability, low temperature, reduced nitrogen, and long-day photoperiods promote rhizome formation, while conditions such as drought, high temperature, and short-day photoperiods favor tillering (McIntyre 1964, 1967; McIntyre and Cessna 1998; McIntyre 2001). However, these patterns are not consistent across species. In *Lotus corniculatus*, for instance, rhizome formation is enhanced under short-day autumn conditions (Kallenbach et al. 2001), whereas exposure to cold suppresses rhizome formation (Moser et al. 1968). Temperature and photoperiod jointly determine elongation capacity. In *Poa pratensis*, long-day and high-temperature conditions promote rhizome elongation (Aamlid 1992; Moser et al. 1968) and in *O. longistaminata*, higher temperatures similarly lead to substantially longer rhizomes (Wang et al. 2024a). Additionally, continuous light accelerates elongation in *C. prolifera* (Chen 1971), and long-day conditions similarly enhance rhizome extension in *Zantedeschia* (Anderson 1970). Drought typically reduces rhizome initiation and elongation in mesic species such as *Festuca arundinacea* (Ma et al. 2020), *Leymus chinensis* (Wang et al. 2019), *Carex lasiocarpa* (Yuan et al. 2017), and *Chrysanthemum morifolium* (Zhang et al. 2022), but can instead enhance clonal spread in stress-adapted taxa, including *Triglochin buchenaui* (Tabot and Adams 2012) and *Leymus secalinus* (Zheng et al. 2021).

The environmental regulation of rhizome branching and shoot emergence is similarly complex. Photoperiodic sensitivity varies markedly across species: in *Nelumbo nucifera*, rhizome branching increases under long, warm days (Masuda et al. 2006), whereas in *Alstroemeria*, long days suppress branching and cooler temperatures promote branch formation (Vonk Noordegraaf 1981). High nitrogen supply, long days, and elevated temperature stimulate rhizome-derived shoot formation in *Cyperus. esculentus* (Garg et al. 1967), whereas *Phragmites australis* exhibits a drought-dependent response, with shoot production increasing after 90 days but becoming suppressed after 120 days (Mingyang et al. 2022).

*Hormonal cross-talk and environmental integration*

In *P. pratensis*, the rhizome-abundant ecotype exhibits elevated CK and a low Auxin/CK ratio, a hormonal balance associated with enhanced rhizome formation (Ran et al. 2023), and exogenous $GA_3$ and IAA similarly promote rhizome formation and upward turning in *L. secalinus* (Li et al. 2022a). Stage-resolved hormonal profiling in *N. nucifera* further shows that CK peaks at the early formation stage, whereas GA, IAA, and SA increase during swelling and ABA accumulates predominantly in late swelling (Li et al. 2006), indicating a shift from early cytokinin enrichment to GA/IAA-associated expansion and ABA-associated maturation. This stage-specific progression highlights that rhizome development relies on the coordinated action of multiple hormone classes. In line with this broader pattern, GA stimulates rhizome formation in *C. morifolium* (Zhang et al. 2022) and promotes elongation in *F. arundinacea* (Ma and Huang 2016; Ma et al. 2016), but suppresses rhizome-derived shoot formation in *C. esculentu* (Garg et al. 1967). ETH likewise enhances rhizome elongation and branching in *S. tuberosum* (Langille 1972). Hormone–environment interactions further modify these processes. In *K. eriantha* (Gesneriaceae), drought or low water availability increases ABA levels, which stimulate ETH production, suppress aerial bud growth, and promote rhizome differentiation (Almeida et al. 2005). By contrast, in *F. arundinacea*, drought suppresses rhizome initiation and growth and is accompanied by increased ABA and soluble sugar accumulation, whereas post-drought recovery involves rises in IAA, CK, and GA that reflect a sequential hormonal rebalancing coordinated with energy metabolism (Ma et al. 2020).

*Synthesis and future directions for in vivo research*

Compared with *in vitro* systems, *in vivo* studies reveal more species-specific and context-dependent outcomes (Table 1b), reflecting the combined action of multiple interacting factors within an integrated physiological framework. Nevertheless, such work has yielded broader, system-level insights into how hormonal cross-talk interfaces with developmental polarity, environmental cues, and carbon allocation to shape rhizome development in whole plants. Despite these advances, most *in vivo* studies rely on endpoint sampling under specific environmental or developmental conditions, restricting our understanding of the dynamic and spatial regulation of rhizome development. In addition, differences in developmental staging, environmental treatments, and sampled tissue types across studies make direct comparisons among species difficult. To address these constraints, new technologies now permit more resolving *in vivo* analyses. Future research should combine controlled environmental manipulations with molecular and imaging analyses in genetically tractable systems, enabling dynamic in situ tracking of hormone distribution, metabolic status, and gene expression (e.g., growth chambers with programmable light/temperature regimes; live reporters such as DR5::GFP, TCSn::GFP, and DII-VENUS; spatial and single-cell transcriptomics; confocal and light-sheet microscopy). Such integrative approaches will help bridge the mechanistic gap between controlled *in vitro* systems and the complex reality of rhizome development in nature.

**Table 1. Qualitative summary of phytohormonal and environmental influences on rhizome development.** (a) Controlled *in vitro* systems; (b) Whole-plant *in vivo* studies. Symbols indicate general trends synthesized across studies: ↑ promoting; ↓ inhibiting; ↑↓ context-dependent or bidirectional; – data not available. Opt-P, Optimal photoperiod; Opt-T, Optimal temperature; LD, long day; Low-T, low temperature.

**(a)**

|  | Initiation | Elongation | Branching | Shooting | Rooting |
|---|---|---|---|---|---|
| Auxin | ↑ | ↑ | ↑ | ↑↓ | ↑↓ |
| Cytokinin | ↑ | ↑ | ↓ | ↑ | ↑↓ |
| Gibberellin | ↑ | ↑ | – | ↑↓ | ↓ |
| Ethylene | ↑ | ↑ | ↑ | ↑↓ | – |
| Jasmonic acid | ↑ | – | – | – | – |
| Abscisic acid | ↑ | – | – | – | – |
| Sucrose | ↑ | ↑ | – | ↑ | ↑ |
| Opt-P | ↑ | – | – | – | – |
| Opt-T | ↑ | – | – | ↑ | – |
| Drought | ↑ | – | – | ↓ | – |

**(b)**

|  | Initiation | Elongation | Branching | Shooting | Rooting |
|---|---|---|---|---|---|
| Auxin | ↑↓ | ↑ | – | – | – |
| Cytokinin | ↑ | ↑ | – | – | – |
| Gibberellin | ↑ | ↑ | – | ↓ | – |
| Ethylene | ↑ | ↑ | ↑ | – | – |
| Jasmonic acid | – | ↑ | – | – | – |
| Abscisic acid | ↑↓ | ↑↓ | – | – | – |
| Nitrogen | ↓ | – | – | ↑ | – |
| LD | ↑↓ | ↑ | ↑↓ | ↑ | – |
| Low-T | ↑↓ | ↓ | ↓↑ | ↓ | – |
| Drought | ↑↓ | ↑↓ | – | ↑↓ | – |

Note: Effects represent qualitative patterns rather than quantitative dose–response relationships. Actual outcomes vary with hormone concentration, stress intensity, species, developmental stage, tissue context, and experimental system. Detailed experimental conditions corresponding to each reported effect are provided in Supplementary Table S1.

## *The molecular mechanisms underlying rhizome initiation and growth*

### *The genetics of rhizome development*

The genetic basis of rhizome initiation and growth has been primarily investigated in economically important crops such as rice (Hu et al. 2003; Li et al. 2022b), sorghum (Kong et al. 2015; Kong et al. 2022; Paterson et al. 1995), and lotus (Huang et al. 2021). These studies have focused on two major developmental aspects: rhizome initiation and subsequent growth traits contributing to rhizome abundance (e.g., rhizome number, branching degree, or spread perimeter). In crops, the presence or absence of rhizomes is generally governed by a polygenic architecture. For instance, Kong et al. (2015) identified at least four loci underlying rhizomatousness between *S. bicolor* and *S. propinquum*, while Hu et al. (2003) reported two dominant-complementary loci, *Rhz2* and *Rhz3*, that regulate rhizome initiation in a cross between cultivated rice (*Oryza sativa*) and wild rice (*O. longistaminata*). Using higher-density markers, Li et al. (2022b) further identified 13 loci associated with rhizome presence in the same species pair and showed that three or more alleles from *O. longistaminata* were necessary for rhizome formation in a recombinant inbred line (RIL) population. Similarly, traits contributing to rhizome abundance have been shown to be polygenically controlled in crops, with key traits including rhizome number (Hu et al. 2003; Huang et al. 2021; Kong et al. 2015; Li et al. 2022b), branching degree (Hu et al. 2003; Li et al. 2022b), and spread perimeter (Larson et al. 2014). However, recent research in *Mimulus* suggests that both rhizome initiation and growth traits underlying abundance (e.g., number, branching, length, width) follow a comparatively simple genetic architecture in crosses between multiple high-altitude perennials and a shared low-altitude perennial (Chen et al. 2025; Coughlan et al. 2021). This contrast may reflect fundamental differences in evolutionary trajectories and ecological strategies across lineages. In crops, rhizomes are often associated with conserved and multifunctional roles such as vegetative reproduction, nutrient storage, and perennial growth, as documented in perennial rice, rhizomatous sorghum, *Miscanthus*, and switchgrass, in which rhizomes function as long-term belowground biomass reserves supporting seasonal regrowth and persistence (Dohleman et al. 2012; Mitros et al. 2020; Paterson et al. 2020; Shanmugam et al. 2025; Silva et al. 2024; Tejera-Nieves and Walker 2023; Tong et al. 2023), roles that are likely underpinned by more layered genetic regulation. By contrast, in *Mimulus*, rhizomes appear to serve primarily as an adaptive trait enabling perenniality in cold, high-elevation environments where vegetative persistence is advantageous, a pattern reflected in both the frequent co-occurrence of rhizomatous growth and perennial life-history strategies in high-altitude taxa and a comparatively less complex genetic architecture underlying rhizome initiation and growth (Chen et al. 2025; Coughlan et al. 2021). Furthermore, although rhizome initiation and subsequent growth traits (e.g. axillary bud fate specification versus rhizome elongation and girth enlargement) are often treated as distinct developmental phases in anatomical and physiological studies (Guo et al. 2021; Yang et al. 2015; Yoshida et al. 2016), QTL mapping often reveals overlapping loci controlling both processes, suggesting shared regulatory mechanisms. For example, *Rhz2* and *Rhz3*, which control rhizome presence in rice, were also found to influence traits such as branching and internode length (Hu et al. 2003), and all five QTLs associated with rhizome number in sorghum were reported to overlap with regions controlling rhizome presence (Kong et al. 2015), collectively pointing to the interpretation that pleiotropic or tightly linked loci can influence both rhizome initiation and subsequent elaboration.

Rhizomes develop from axillary buds, and accumulating evidence suggests a close genetic relationship between axillary bud regulation and rhizome development. For example, Paterson et al. (1995) demonstrated that one QTL influenced the number of available axillary buds, while several others determined the fate of these buds, that is, whether they differentiate into rhizomes or tillers, in crosses between wild and cultivated *Sorghum. bicolor*. Similarly, Kong et al. (2015) found that some QTLs associated with rhizome occurrence in *S. bicolor* × *Sorghum. propinquum* overlapped with QTLs controlling tiller number and axillary branching, suggesting that rhizomatousness and vegetative branching may share common developmental pathways. This pattern was further supported by Kong et al. (2022), who identified overlapping QTLs for rhizomatousness, tillering, and branching in *S. bicolor* × *S. halepense* crosses. In addition to these shared regulatory axes, Hu et al. (2003) observed that QTLs controlling rhizome initiation and growth traits underlying abundance in *O. longistaminata* largely corresponded to homologous regions in the *S. propinquum* genome, indicating that certain loci involved in rhizome development may be conserved across Poaceae lineages. Although causal genes have yet to be fully identified, these findings collectively provide important insights into the genetic mechanisms underlying rhizome development and lay the groundwork for future gene-level discovery through integrative genetic and genomic approaches.

## *Gene regulation networks of rhizomes based on multiple expression data*

Despite the lack of identified causal genes directly responsible for rhizome initiation and proliferation in natural plant populations, extensive transcriptomic investigations have been conducted across a wide array of rhizomatous species to uncover candidate genes potentially involved in rhizome development. These high-throughput gene expression studies span economically and ecologically important species such as *O. longistaminata* (He et al. 2014; Hu et al. 2011; Wang et al. 2024a; Zhang et al. 2015), *S. halepense* and *S. propinquum* (Jang et al. 2006; Zhang et al. 2014), *N. nucifera* (Cheng et al. 2013; Ming et al. 2013; Yang et al. 2015), *A. lancea* (Huang et al. 2016), *Miscanthus lutarioriparius* (Hu et al. 2017), *Trifolium ambiguum* (Meng et al. 2021), *L. chinensis* (Gao et al. 2023), as well as *Z. officinale* and *Curcuma. longa* (Koo et al. 2013). A convergent finding across these studies is the recurrent involvement of phytohormone-related genes in rhizome development, particularly those involved in Aux, CK, ETH, GA, and ABA signaling pathways. For example, transcriptomic analyses comparing stolon tips (at the rhizome induction stage) with other stages of rhizome development (e.g., swelling stages) in *N. nucifera* revealed elevated expression of auxin- and ethylene-responsive genes, suggesting a role for these pathways in rhizome initiation (Cheng et al. 2013). Most transcriptomic studies, however, have focused on differential gene expression between rhizome tissues and non-rhizome tissues (e.g., leaves). In *O. longistaminata*, genes involved in Aux, GA, and ETH signaling showed higher expression in rhizomes than in leaves, roots, and stems (He et al. 2014), whereas in *A. lancea*, cytokinin- and auxin-related genes were preferentially upregulated in rhizome tissues (Huang et al. 2016). Similar hormone-related expression patterns have also been described across the rhizomatous taxa listed above, reinforcing the central role of hormone-mediated regulatory networks in rhizome development across diverse lineages.

Moreover, many genes associated with biotic and abiotic stress responses, such as heat shock proteins, have been reported to be highly expressed in rhizome tissues across multiple species (He et al. 2014; Huang et al. 2016; Ming et al. 2013). In parallel, genes involved in energy metabolism, particularly those related to starch and sucrose pathways, also exhibit elevated expression in rhizomes (Cheng et al. 2013; Hu et al. 2017; Huang et al. 2016; Meng et al. 2021), consistent with the established roles of rhizomes in nutrient storage and overwintering. Recent omics work on rhizomatous tissues also commonly reports elevated expression of lignin-biosynthetic genes, indicating that enhanced cell-wall reinforcement is another recurring feature of rhizome development (Ma et al. 2024; Meng et al. 2021; Yang et al. 2019; Zhengyan et al. 2023). In addition, a suite of transcription factors, including members of the AP2/ERF, bHLH, bZIP, WRKY, and MADS-box families, are commonly expressed in rhizomatous tissues across diverse taxa (Fig. 2a) (Cheng et al. 2013; He et al. 2014; Hu et al. 2011; Huang et al. 2016; Koo et al. 2013; Meng et al. 2021; Ming et al. 2013; Ruan et al. 2021; Yang et al. 2015; Zhang et al. 2014). These transcription factors are broadly implicated in plant hormone signaling and environmental stress responses, reinforcing the view that rhizome development is tightly coordinated by integrated hormonal and stress-response regulatory networks.

### *Genes functionally validated to influence rhizome development*

To date, only a few genes have been functionally validated to influence rhizome development. One such gene is *BLADE-ON-PETIOLE* (*BOP*), a conserved regulator of proximal–distal leaf patterning (Hepworth et al. 2005; Toriba et al. 2019) rather than a rhizome-specific factor. In the rhizomes of *O. longistaminata*, high *BOP* activity maintained by miR156 suppresses leaf blade formation, yielding compact, sheath-dominated leaves; *bop* loss-of-function mutants instead produce ectopic blades on rhizome leaves, which reduces tip stiffness and impairs penetration through soil (Toriba et al. 2020). Thus, *BOP* primarily controls leaf identity, but this modulation of rhizome leaf morphology has important consequences for underground growth mechanics. Two additional genes, *CmRH56* and its downstream target *CmGA2ox6*, have been identified in chrysanthemum (*C. morifolium*) (Zhang et al. 2022). *CmRH56*, a DEAD-box RNA helicase specifically expressed in the rhizome shoot apex, regulates rhizome outgrowth under drought stress: RNAi knockdown lines produce fewer rhizomes, whereas overexpression accelerates and increases rhizome formation. Mechanistically, *CmRH56* represses the GA-catabolic gene *CmGA2ox6*. Silencing *CmGA2ox6* leads to a marked increase in rhizome number, indicating that GA availability promotes rhizome outgrowth. Importantly, however, current evidence does not distinguish whether GA signaling specifically biases axillary buds toward a rhizome identity, or more generally enhances lateral bud outgrowth that yields both aerial and subterranean stems. Despite this uncertainty, these gene-based studies provide initial insights into the molecular mechanisms underlying rhizome development.

### *Synthesis and future directions for genetic and regulatory studies of rhizomes*

Although genetic mapping, expression profiling, and limited functional studies have begun to identify loci and pathways involved in rhizome initiation and growth, the mechanisms that specify rhizome identity and govern its subsequent growth remain poorly understood. In particular, it is still unclear how hormonal cues, environmental signals, and developmental regulators interact to direct an axillary bud toward a rhizome rather than an aerial shoot. Addressing these questions will require moving from correlation to causation through precise genetic dissection of the bud-to-rhizome transition and its downstream developmental programs. Single-cell and spatial transcriptomics, high-resolution fine mapping in near isogenic lines, forward genetic screens such as EMS mutagenesis, *cis*-regulatory profiling using methods such as ATAC seq or CUT and RUN, and CRISPR based perturbation of candidate regulators together represent the most promising strategies for identifying the core regulatory switches that initiate and sustain rhizome development.

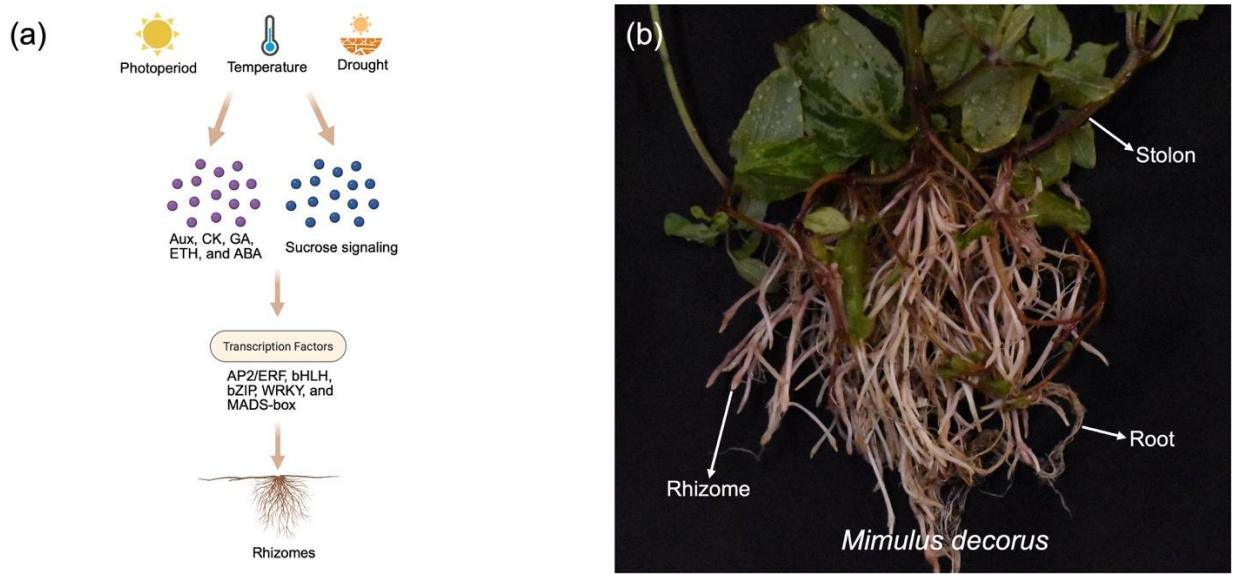

**Fig. 2. Molecular and morphological basis of rhizome development.** (a) Conceptual model illustrating how rhizome development is modulated. Aux: auxin; CK: cytokinin; GA: gibberellin; ETH: ethylene; ABA: abscisic acid; (b) Photograph of *Mimulus decorus* showing clear morphological differentiation between rhizomes (horizontal underground stems), stolons (horizontal aboveground stems), and roots.

### *Evolutionary roles and adaptive significance of rhizomes*

Rhizomes play a pivotal role in the evolutionary history of vascular plants, both as ancestral traits and as drivers of ecological diversification. In monocots, the ancestral condition is hypothesized to have been aquatic with a rhizomatous growth form, which enabled plants to anchor in shallow water while adjusting to fluctuations in water and sediment levels, thereby improving gas exchange efficiency (Howard et al. 2019). Some studies further propose that the rhizome represents a morphological precursor to the upright stem (McDowell and Gang 2012), and environmental factors such as temperature have likely served as key selective pressures shaping rhizome diversity (Howard et al. 2019). The evolutionary significance of rhizomes becomes particularly evident in the context of plant invasiveness, where rapid evolution is frequently observed. For instance, *S. halepense*, native to western Asia, exhibits a strong association between its rhizomatous growth and its acquired cold tolerance following introduction to North America (Paterson et al. 2020). Similarly, *Artemisia vulgaris*, a European native, has evolved extensive rhizome networks that contribute to its invasiveness across the Northeastern and Mid-Atlantic United States (Barney et al. 2009). Rhizomes confer adaptive advantages under extreme conditions such as drought and frost, primarily by serving as energy storage organs that support regrowth after stress or dormancy (McDowell and Gang 2012). Additionally, rhizomes enhance ecosystem engineering by improving soil stabilization, which may have contributed to the successful terrestrial radiation of early vascular plants (Xue et al. 2016).

Together, these examples indicate that rhizomes represent a repeatedly favored evolutionary strategy for persistence, resource buffering, and environmental tracking across diverse lineages. Several major evolutionary questions remain unresolved, including the frequency of independent origins of rhizomatous growth, the extent to which convergent ecological functions rely on shared developmental and genetic modules, and the ways in which rhizome evolution alters life-history trade-offs with reproduction and dispersal. Emerging approaches such as comparative single-cell atlases, CRISPR-based functional assays, and high-throughput phenotyping under controlled conditions now offer powerful tools for addressing these gaps. Future progress will benefit from integrating phylogenetic comparative analyses with developmental genetics, genomics, and experimental tests of performance across heterogeneous environments. This combined framework will help clarify when, why, and how rhizomes evolve, and help explain their persistent prevalence as an architectural strategy across plant lineages.

## *Mimulus as a model system for rhizome biology*

While extensive research has been conducted on rhizomes across diverse plant taxa, key mechanistic insights remain limited. Real progress on the questions and approaches outlined above will require a system that integrates natural variation in rhizome traits with strong genetic and genomic tractability, making the identification of an appropriate model organism essential for advancing rhizome research.

Most existing model systems, however, fall short of meeting these requirements. Classical genetic models such as *Arabidopsis thaliana*, *Solanum lycopersicum*, *Brachypodium distachyon*, *Setaria viridis*, and *Populus trichocarpa* do not produce rhizomes, which fundamentally limits their utility for studying rhizome initiation or underground stem development. In rice, *O. sativa* forms rhizomes almost exclusively through introgression from its rhizomatous wild relative *O. longistaminata*, a perennial species characterized by long generation times (Cissé and Khouma 2016) and low transformability (Shimizu-Sato et al. 2020), which limits the suitability of this system for mechanistic dissection of rhizome biology. Against this backdrop, *Mimulus* (monkeyflowers) represents a particularly promising alternative. It possesses traits absent from classical model systems such as *Arabidopsis*, including diverse floral pigmentation patterns, corolla tube formation, and robust rhizomatous growth (Fig. 2b) (Yuan 2019), and species within this genus share additional attributes that facilitate genetic and evolutionary research, including small genome sizes, short generation times, high fecundity, and extensive natural variation in mating systems, life history traits, and edaphic preferences (Wu et al. 2008; Yuan 2019). Notably, at least four species complexes, namely *M. aurantiacus*, *M. guttatus*, *M. luteus*, and *M. lewisii*, are amenable to *Agrobacterium*-mediated transformation (Yuan 2019), facilitating molecular genetic studies via transgenic perturbation. A side-by-side comparison of these systems (Table 2) further underscores the unique suitability of *Mimulus* for advancing rhizome biology. This suitability has supported recent breakthroughs in the genetic control of floral pigmentation (Ding et al. 2020a; LaFountain et al. 2023; Sagawa et al. 2016; Stanley et al. 2020) and floral morphology (Chen et al. 2022; Ding et al. 2021; Ding et al. 2017; Ding et al. 2020b; Yuan et al. 2013a), based on approaches such as EMS mutagenesis and reverse genetics. Moreover, several causal genes underlying natural variation in flower color have been identified and functionally validated (Liang et al. 2023; Liang et al. 2022; Yuan et al. 2016; Yuan et al. 2013b), advancing our understanding of the molecular mechanisms and evolutionary processes shaping phenotypic diversity in natural populations. Collectively, these attributes position *Mimulus* as a powerful and underutilized model system with exceptional potential for uncovering the genetic and developmental basis of rhizome traits.

Building on recent QTL mapping efforts in the *M. guttatus* species complex that revealed the genetic architecture of repeated life-history divergence associated with high-altitude adaptation (Chen et al. 2025), future research should aim to dissect the genetic basis of rhizome variation in this system with greater precision. By integrating complementary approaches, including near-isogenic line development, fine-scale genetic mapping, multi-omics network reconstruction, and transgenic perturbation, it will be possible to identify key regulatory loci and decode the molecular mechanisms underlying rhizome initiation and elaboration. Such efforts will not only deepen understanding of how rhizomes evolve and function within ecological and evolutionary contexts but will also establish *Mimulus* as a powerful and genetically tractable model for rhizome biology. Ultimately, this line of research has the potential to bridge a long-standing gap between developmental genetics and ecological adaptation, shedding light on the genetic strategies that perennial plants use to persist across heterogeneous environments.

**Table 2. Comparison of major plant model systems relevant to dissecting rhizome biology.** Summary reflects consolidated information from published studies on model-system life cycles, genome characteristics, and transformation methods (representative references cited in the main text).

| Model system | Rhizomes | Generation time | Genome size | Natural variation in rhizomes | Transformation efficiency |
|---|---|---|---|---|---|
| *Arabidopsis thaliana* | × | ~6 weeks | ~135 Mb | × | √ |
| *Oryza sativa* (rice) | ✗ in *O. sativa*; ✓ in wild relative | ~4 months | ~400 Mb | ✗ (via *O. longistaminata* introgression) | ✓ *O. sativa*; ✗ *O. longistaminata* |
| *Solanum lycopersicum* (tomato) | ✗ | ~3 months | ~900 Mb | ✗ | √ |
| *Brachypodium distachyon* | ✗ | ~8 weeks | ~272 Mb | ✗ | √ |
| *Setaria viridis* | ✗ | ~6 weeks | ~510 Mb | ✗ | √ |
| *Populus trichocarpa* (Poplar) | ✗ | ~6 months (early-flowering lines) | ~480 Mb | ✗ | √ |
| *Mimulus* (monkeyflowers) | √ | ~6 weeks | ~430 Mb | √ extensive | √ |

# Acknowledgements

Figure cartoons (Fig. 1a, 2a) were created using BioRender.com.

# Funding

This work was supported by the Brown Endowed Fellowship at Yale University, which funded H.F.C. for one year, and by the National Institutes of Health (NIH R35GM150907) awarded to J.M.C.

# Author's contribution

H.F.C. led the conceptual development and writing of the manuscript. J.M.C. provided intellectual oversight, critical feedback, and contributed to manuscript revision.

# Availability of data and materials

No new datasets were generated or analyzed for this study.

# Declarations

**Conflict of interest**

The author declares no conflict of interest.

Sagawa JM, Stanley LE, LaFountain AM, Frank HA, Liu C, Yuan YW (2016) An R2R3-MYB transcription factor regulates carotenoid pigmentation in *Mimulus lewisii* flowers. New Phytol 209:1049-1057.

Shahriari AG, Bagheri A, Sharifi A, Moshtaghi N (2012) Efficient regeneration of 'Caralis' *Alstroemeria* cultivar from rhizome explants. Not Sci Biol 4:86-90.

Shanmugam V, Tyagi VC, Rajendran G, Chimmili SR, Swarnaraj AK, Arulanandam M, Kumar V, Peramaiyan P, Murugaiyan V, Sundaram RM (2025) Perennial rice–An alternative to the 'one-sow, one-harvest'rice production: Benefits, challenges, and future prospects. Farming Syst 3:100137.

Shimasaki K (1993) Effect of auxin and ethylene on rhizome formation from shoot cultures of *Cymbidium kanran* Makino. Plant Tissue Cult Lett 10:156-159.

Shimasaki K (1995) Interactive effects between cytokinin and ethephon on shoot formation in rhizome cultures of *Cymbidium kanran* Makino. Plant Tissue Cult Lett 12:27-33.

Shimasaki K, Uemoto S (1990) Micropropagation of a terrestrial *Cymbidium* species using rhizomes developed from seeds and pseudobulbs. Plant Cell Tissue Organ Cult 22:237-244.

Shimasaki K, Uemoto S (1991) Rhizome induction and plantlet regeneration of *Cymbidium goeringii* from flower bud cultures in vitro. Plant Cell Tissue Organ Cult 25:49-52.

Shimizu-Sato S, Tsuda K, Nosaka-Takahashi M, Suzuki T, Ono S, Ta KN, Yoshida Y, Nonomura K-I, Sato Y (2020) *Agrobacterium*-mediated genetic transformation of wild *Oryza* species using immature embryos. Rice 13:33.

Silva GC, Yu J, Herndon L, Samuelson S, Rajan N, Bagavathiannan M (2024) Evaluation of organic options for Johnsongrass (*Sorghum halepense*) control during winter fallow. Weed Sci 72:275-283.

Stanley LE, Ding B, Sun W, Mou F, Hill C, Chen S, Yuan Y-W (2020) A tetratricopeptide repeat protein regulates carotenoid biosynthesis and chromoplast development in monkeyflowers (*Mimulus*). Plant Cell 32:1536-1555.

Subramani V, Kamaraj M, Ramachandran B, Jeyakumar JJ (2014) Effect of different growth regulators on in-vitro regeneration of rhizome and leaf explants of *Acorus calamus* L. Int J Pharm Res Rev 3:1-6.

Tabot P, Adams J (2012) Morphological and physiological responses of *Triglochin buchenaui* Köcke, Mering & Kadereit to various combinations of water and salinity: implications for resilience to climate change. Wetl Ecol Manag 20:373-388.

Tejera-Nieves M, Walker BJ (2023) A 30% reduction in switchgrass rhizome reserves did not decrease biomass yield. GCB Bioenergy 15:1329-1338.

Tikendra L, Sushma O, Amom T, Devi NA, Paonam S, Bidyananda N, Potshangbam AM, Dey A, Devi RS, Nongdam P (2022) Genetic clonal fidelity assessment of rhizome-derived micropropagated Acorus calamus L.–A medicinally important plant by random amplified polymorphic DNA and inter-simple sequence repeat markers. Pharmacogn Mag 18:207-215.
22

**Table S1. Experimental conditions associated with reported hormonal or environmental effects on rhizome traits.** Opt-P, Optimal photoperiod; Opt-T, Optimal temperature; LD, long day; Low-T, low temperature.

| Hormone / Factor | Species | System (in vivo / in vitro) | Tissue | Treatment (concentration / stress level) | Duration | Reported effect | Citation |
|---|---|---|---|---|---|---|---|
| Auxin (IBA; NAA; IAA) | *Cymbidium forrestii* | in vitro | rhizome | 0.1–5 mg/L | 12 weeks | promotes rhizome initiation | Paek and Yeung 1991 |
| Auxin (NAA) | *Cymbidium goeringii* | in vitro | Apical flower buds | 1 mg/L | 8 weeks | promotes rhizome initiation | Shimasaki and Uemoto 1991 |
| Auxin (IAA) | *Leymus secalinus* | in vivo | whole plants | 0.005 mg/L | 14 days | promotes rhizome initiation | Li et al. 2022a |
| Auxin (NAA) plus Cytokinin (6-BA) | *Gastrodia elata* | in vitro | juvenile rhizome | 1–2 mg/L (NAA); 2 mg/L (BA) | 8 weeks | promotes rhizome elongation | Hsieh et al. 2022 |
| Auxin (IAA; IBA; NAA) | *Cymbidium aloifolium* | in vitro | rhizomes | 0.1–5 mg/L | 8 weeks | promotes rhizome branching | Nayak et al. 1998 |
| Auxin (NAA) | *Geodorum densiflorum* | in vitro | protocorm | 1 mg/L | 3 months | promotes rhizome branching | Roy and Banerjee 2002 |
| Auxin (IBA; NAA; IAA) | *Cymbidium forrestii* | in vitro | rhizome | 0.1–5 mg/L | 12 weeks | promotes rhizome branching | Paek and Yeung 1991 |
| Auxin (NAA) | *Geodorum densiflorum* | in vitro | protocorm | 0.5-4 mg/L | 2–6 months | promotes rhizome shooting | Roy and Banerjee 2002 |
| Auxin (NAA) | *Alstroemeria* | in vitro | rhizomes | 0.2 mg/L | 9 weeks | promotes rhizome shooting | Khaleghi et al. 2008 |
| Auxin (NAA) | *Alstroemeria* | in vitro | rhizome buds | 0.2 mg/L | 12 weeks | promotes rhizome shooting | Hamidoghli et al. 2007 |
| Auxin (IBA; IAA) | *Acorus calamus* | in vitro | rhizomes | 0.8 mg/L | 3 weeks | inhibits rhizome shooting | Tikendra et al. 2022 |
| Auxin (NAA) | *Ruppia maritima* | in vitro | terminal rhizome | 1 mg/L | 12 weeks | inhibits rhizome rooting | Koch and Durako 1991 |

| Hormone | Species | Condition | Explant | Concentration | Duration | Effect | Reference |
|---|---|---|---|---|---|---|---|
| Auxin (NAA) | *Alstroemeria* | in vitro | rhizome segments | 0.2 mg/L | 9 weeks | promotes rhizome rooting | Khaleghi et al. 2008 |
| Auxin (NAA) | *Alstroemeria* | in vitro | rhizome buds | 0.2–1 mg/L | 12 weeks | promotes rhizome rooting | Hamidoghli et al. 2007 |
| Auxin (NAA) | *Alstroemeria* | in vitro | rhizome | 1 µM | 7 weeks | promotes rhizome rooting | Kristiansen et al. 1999 |
| Auxin (IBA) | *Podophyllum hexandrum* | in vitro | rhizomes | 100 µM | 24 h | promotes rhizome rooting | Nadeem et al. 2000 |
| Auxin (NAA; IBA) | *Posidonia oceanica* | in vitro | rhizomes | 5 mg/L | 24 h | promotes rhizome rooting | Balestri and Lardicci 2006 |
| Auxin (IBA) | *Scopolia parviflora* | in vitro | rhizome | 2.46 µM | 4 weeks | promotes rhizome rooting | Kang et al. 2004 |
| Auxin (NAA) | *Valeriana jatamansi* | in vitro | rhizome | 1–1.5 mg/L | 8 weeks | promotes rhizome rooting | Nazir et al. 2022 |
| Cytokinin (6-BA) | *Cyperus serotinus* | in vitro | tubers | 0.1–10 mg/L | 14 days | promotes rhizome initiation | Omokawa et al. 1992 |
| Cytokinin (6-BA; kinetin) | *Kaempferia galang* | in vitro | rhizome | 6-BA: 13.2 µM; kinetin: 4.6 µM | 120 days | promotes rhizome shooting | Vincent et al. 1992 |
| Cytokinin (6-BA) | *Curcuma aromatica* | in vitro | shoots | 1–5 mg/L | 30 days | promotes rhizome initiation | Nayak 2000 |
| Cytokinin (6-BA) | *Zingiber officinale* | in vitro | shootlets | 6–9 mg/L | 10 weeks | promotes rhizome initiation | Abbas et al. 2014 |
| Cytokinin (6-BA; 2iP; Kinetin) | *Cymbidium forrestii* | in vitro | rhizome | 0.5–10 mg/L | 12 weeks | inhibits rhizome branching | Paek and Yeung 1991 |
| Cytokinin (Zeatin; 6-BA) | *Cymbiaium kanran* | in vitro | rhizome apical segments | 0.01–10 µM | 8 weeks | inhibits rhizome branching | Shimasaki 1995 |

| Hormone | Species | Condition | Explant | Concentration | Duration | Effect | Reference |
|---|---|---|---|---|---|---|---|
| Cytokinin (6-BA) | *Geodorum densiflorum* | in vitro | protocorm | 0.5–8 mg/L | 3 months | promotes rhizome shooting | Roy and Banerjee 2002 |
| Cytokinin (6-BA) | *Geodorum densiflorum* | in vitro | protocorm | 0.5–8 mg/L | 3 months | inhibits rhizome rooting | Roy and Banerjee 2002 |
| Cytokinin (Kinetin; 6-BA; 2iP; Zeatin; thidiazuron) | *Ruppia maritima* | in vitro | terminal rhizome segments | 5–20 mg/L (BAP, 2iP, and zeatin); $10^{-9}$ (thidiazuron) | 12 weeks | inhibits rhizome rooting | Koch and Durako 1991 |
| Cytokinin (6-BA) | *Gastrodia elata* | in vitro | juvenile rhizome | 1 mg/L | 8 weeks | promotes rhizome elongation | Hsieh et al. 2022 |
| Cytokinin (6-BA; Kinetin) | *Cymbidium aloifolium* | in vitro | rhizomes | 0.25–1 mg/L | 8 weeks | promotes rhizome shooting | Nayak et al. 1998 |
| Cytokinin (Zeatin; 6-BA) | *Cymbiaium kanran* | in vitro | rhizome apical segments | 0.01–10 µM | 8 weeks | promotes rhizome shooting | Shimasaki 1995 |
| Cytokinin (6-BA) | *Alstroemeria* | in vitro | rhizomes | 0.5–2.5 mg/L | 9 weeks | promotes rhizome shooting | Khaleghi et al. 2008 |
| Cytokinin (6-BA) | *Cymbidium kanran* | in vitro | rhizome apex | 0.1–10 mg/L | 4 weeks | promotes rhizome shooting | Shimasaki and Uemoto 1990 |
| Cytokinin (6-BA; TDZ) | *Acorus calamus* | in vitro | rhizomes | 0.8–2.4 mg/L | 3 weeks | promotes rhizome shooting | Tikendra et al. 2022 |
| Cytokinin (6-BA、2iP) | *Alstroemeria* | in vitro | Rhizomes | 0.5–2 mg/L | 6–8 weeks | inhibits rhizome rooting | Gabryszewska and Hempel 1984 |
| Cytokinin (6-BA) | *Alstroemeria* | in vitro | rhizome | 10 µM | 7 weeks | inhibits rhizome rooting | Kristiansen et al. 1999 |
| Auxin (NAA) plus Cytokinin (6-BA) | *Cymbidium goeringii* | in vitro | Apical flower buds | 0.1 mg/L 6-BA; 10 mg/L NAA | 8 weeks | promotes rhizome initiation | Shimasaki and Uemoto 1991 |

| Hormone | Species | Condition | Tissue | Concentration | Duration | Effect | Reference |
|---|---|---|---|---|---|---|---|
| Auxin (NAA) plus Cytokinin (6-BA) | *Bambusa bambos* | in vitro | shoots | 2.5 μM 6-BA; 50 μM NAA | 3–4 weeks | promotes rhizome initiation | Kapoor and Rao 2006 |
| Auxin (NAA) plus Cytokinin (6-BA) | *Geodorum densiflorum* | in vitro | protocorm | 4 mg/L 6-BA; 1 mg/L NAA | 3 months | promotes rhizome initiation | Roy and Banerjee 2002 |
| Auxin (NAA) plus Cytokinin (6-BA) | *Geodorum densiflorum* | in vitro | protocorm | 0.5–4 mg/L 6-BA; 1 mg/L NAA | 3 months | Inhibits rhizome branching | Roy and Banerjee 2002 |
| Auxin (NAA) plus Cytokinin (zeatin) | *Zingiber officinale* | in vitro | shoots | 2.5 μM NAA; 10 μM zeatin | 14 weeks | promotes rhizome initiation | Zahid et al. 2021 |
| Auxin (NAA) plus Cytokinin (6-BA) | *Alstroemeria* | in vitro | rhizome tips | 1 mg/L 6-BA; 0.2 mg/L NAA | 12 weeks | promotes rhizome shooting | Shahriari et al. 2012 |
| Auxin (NAA) plus Cytokinin (6-BA) | *Valeriana jatamansi* | in vitro | rhizomes | 1–1.5 mg/L BAP; 0.5–1.5 mg/L NAA | 5–6 weeks | promotes rhizome rooting | Nazir et al. 2022 |
| gibberellic acid (GA$_3$) | *Bambusa bambos* | in vitro | shoots | 0.1 μM | 3–4 weeks | promotes rhizome initiation | Kapoor and Rao 2006 |
| gibberellic acid (GA$_3$) | potato | in vitro | etiolated sprout segments | 0.2–2 mg/L | 3 weeks | promotes rhizome initiation | Escalante and Langille 1995 |
| gibberellic acid (GA$_3$) | *Leymus secalinus* | in vivo | whole plants | 0.2 mg/L | 14 days | promotes rhizome initiation | Li et al. 2022a |
| gibberellic acid (GA$_3$) | *Chrysanthemum morifolium* | in vivo | whole plants | 100 μM | 50 days (every 10 days treatment) | promotes rhizome initiation | Zhang et al. 2022 |
| gibberellic acid (GA$_3$) | *Caulerpa prolifera* | in vitro | thalli | 1.67 μg/L | 14-25 days | promotes rhizome elongation | Jacobs and Davis 1983 |
| gibberellic acid (GA$_3$) | potato | in vitro | potato | 100 mg/L | 10 days | promotes rhizome | Escalante and |

| Hormone | Species | Condition | Tissue | Concentration | Duration | Effect | Reference |
|---|---|---|---|---|---|---|---|
| | | | | | | elongation | Langille 1998 |
| gibberellic acid (GA₃) | *Festuca arundinacea* | in vivo | whole plants | 10 μM | 12 days | promotes rhizome elongation | Ma and Huang 2016 |
| gibberellic acid (GA₃) | *Festuca arundinacea* | in vivo | whole plants | 10 μM | 12 days | promotes rhizome elongation | Ma et al. 2016 |
| gibberellic acid (GA₃) | *Cyperus esculentu* | in vivo | whole plants | 1000 ppm | 28 days | inhibits rhizome shooting | Garg et al. 1967 |
| gibberellic acid (GA₃) | *Rheum rhabarbarum* | in vitro | rhizome | 5 and 50 mg/L | 1 week | promotes rhizome shooting | Rayirath et al. 2009 |
| gibberellic acid (GA₃) | *Agropyron repens* | in vitro | rhizomes | 10 mg/L | 14 days | inhibits rhizome shooting | Rogan and Smith 1976 |
| gibberellic acid (GA₃) | *Zantedeschia* | in vitro | rhizome | 150 mg/L | 20 minutes | promotes rhizome shooting | Kozłowska et al. 2007 |
| gibberellic acid (GA₃) | *Convolvulus sepium* | in vitro | rhizome | 0.02–1 mg mg/L | 1 week | inhibits rhizome rooting | Wells and Riopel 1972 |
| Ethylene (Ethephon) | *Rheum rhabarbarum* | in vitro | shoot clumps | 1–50 mg/L | 6 weeks | promotes rhizome initiation | Rayirath et al. 2011 |
| Ethylene (Ethephon) | *Cymbidium kanran* | in vitro | shoots | 1–100mg/L | 7 weeks | promotes rhizome initiation | SHIMASAKI 1993 |
| Ethylene (Ethephon) | *Rheum rhabarbarum* | in vitro | shoot clumps | 1 mg/L | 6 weeks | promotes rhizome elongation | Rayirath et al. 2011 |
| Ethylene (Ethephon) | *Rheum rhabarbarum* | in vitro | shoot clumps | 1 mg/L | 6 weeks | promotes rhizome branching | Rayirath et al. 2011 |
| Ethylene (Ethephon) | *Cymbiaium kanran* | in vitro | rhizome apical segments | 10 μM | 8 weeks | promotes rhizome branching | Shimasaki 1995 |
| Ethylene (Ethephon) | *Cymbiaium kanran* | in vitro | rhizome apical | 10 μM | 8 weeks | inhibits rhizome shooting | Shimasaki 1995 |

| Hormone | Species | Condition | Explant | Concentration | Duration | Effect | Reference |
|---|---|---|---|---|---|---|---|
| Ethylene (Ethephon) | *Cymbidium kanran* | in vitro | apical rhizome segments | 10 mg/L | every 2 weeks, up to 4 months | inhibits rhizome shooting | Ogura-Tsujita and Okubo 2006 |
| Ethylene (Ethrel) | *Kohleria eriantha* | in vitro | rhizome | $10^{-3}$ M | 10 days | inhibits rhizome shooting | Almeida et al. 2005 |
| Ethylene (ethephon) | *Zingiber officinale* | in vitro | rhizomes | 750 ppm | 10 minutes | promotes rhizome shooting | Furutani et al. 1985 |
| Ethylene (Ethrel) | *Solanum tuberosum* | in vivo | whole plants | 10-100 ppm | 6 weeks | promotes rhizome elongation | Langille 1972 |
| Ethylene (ethephon) | *Solanum tuberosum* | in vio | whole plants | 50-1000 ppm | 6 weeks | promotes rhizome branching | Langille 1972 |
| Jasmonic acid | *Rheum rhabarbarum* | in vitro | shoot clumps | 10 ng/L–1 µg/L | 6 weeks | promotes rhizome initiation | Rayirath et al. 2011 |
| Abscisic acid | *Cnidium officinale* | in vitro | micro shoots | 0.5 mg/L–1.0 mg/L | 8 weeks | promotes rhizome initiation | Kim et al. 2022 |
| Sucrose | *Bambusa bambos* | in vitro | shoots | 5% | 3–4 weeks | promotes rhizome initiation | Kapoor and Rao 2006 |
| Sucrose | *Kaempferia parviflora* | in vitro | rhizome buds | 6% | 9 weeks | promotes rhizome initiation | Labrooy et al. 2020 |
| Sucrose | *Zingiber officinale* | in vitro | shoots | 45 g/L–90 g/L | 12 weeks | promotes rhizome initiation | Zahid et al. 2021 |
| Sucrose | *Zingiber officinale* | in vitro | plantlets | 30–100 g/L | 12 weeks | promotes rhizome initiation | Gezahegn et al. 2024 |
| Sucrose | *Oryza longistaminata* | in vitro | Seedlings | 60–120 g/L | 3 months | promotes rhizome initiation | Fan et al. 2022 |
| Sucrose | *Bambusa bambos* | in vitro | shoots | 5% | 3–4 weeks | promotes rhizome elongation | Kapoor and Rao 2006 |
| Sucrose | *Acorus calamus* | in vitro | shoots | 2%–6% | 6 weeks | promotes rhizome | Subramani et al. 2014 |
| Sucrose | *Oryza longistaminata* | in vitro | Seedlings | 60–100 g/L | 3 months | promotes rhizome elongation | Fan et al. 2022 |
| Sucrose | *Cymbidium forrestii* | in vitro | rhizome | 3–7% | 12 weeks | promotes rhizome shooting | Paek and Yeung 1991 |
| Sucrose | *Zingiber officinale* | in vitro | shoots | 45–60 g/L | 12 weeks | promotes rhizome shooting | Zahid et al. 2021 |
| Sucrose | *Zingiber officinale* | in vitro | shoots | 45–60 g/L | 12 weeks | promotes rhizome rooting | Zahid et al. 2021 |
| Sucrose | *Alstroemeria* | in vitro | rhizome | 20–60 g/L | 8 weeks | promotes rhizome rooting | Gabryszewska 1996 |
| Nitrogen | *Agropyron repens* | In vivo | Whole plants | 2.1 ppm | Transplanting → 6th leaf | promotes rhizome initiation | McIntyre 1964 |
| Nitrogen | *Cyperus esculentus* | In vivo | Whole plants | Hoagland solutions of 1/32–1/2 | 28 days | promotes rhizome shooting | Garg et al. 1967 |
| Opt-P | *Curcuma aromatica* | in vitro | shoots | 8 h/day | 30 days | promotes rhizome initiation | Nayak 2000 |
| Opt-P | *Zingiber officinale* | in vitro | shoots | 8-24h/day | 8 weeks | promotes rhizome initiation | Rout et al. 2001 |
| Photoperiod | *Oryza longistaminata* | in vitro | rhizome | continuous light | 2 weeks | inhibits rhizome initiation | Yoshida et al. 2016 |
| Opt-P | *Cyperus rotundus* | in vitro | tubers | 25°C–35°C | 10 days | promotes rhizome initiation | Maria De Fatima and Valio 1976 |
| Opt-P | *Agropyron repens* | in vitro | rhizome | 13–23°C | 30 days | promotes rhizome shooting | Leakey et al. 1978 |
| LD | *Agropyron repens* | in vivo | Whole plants | 18 h | Transplanting → 6th leaf | promotes rhizome initiation | McIntyre 1967 |
| LD | *Caulerpa prolifera* | in vivo | whole plants | continuous light | 8 days | promotes rhizome elongation | CHEN 1971 |
| LD | *Trientalis boreal* | in vivo | whole plants | 8–17h | 4 months | promotes rhizome elongation | Anderson 1970 |
| LD | *Poa pratensis* | in vivo | whole plants | - | 9 weeks | promotes rhizome elongation | Aamlid 1992 |
| LD | *Poa pratensis* | in vivo | whole plants | 16–18 h | 7–8 weeks | promotes rhizome elongation | Moser et al. 1968 |
| LD | *Nelumbo nucifera* | in vivo | whole plants | Summer long-day | 2 months | promotes rhizome branching | Masuda et al. 2006 |
| LD | *Alstroemeria* | in vivo | whole plants | 12–16 h | 2–11 weeks | inhibits rhizome branching | Vonk Noordegraaf 1981 |
| LD | *Cyperus esculentus* | in vivo | whole plants | 15½ h | 28 days | promotes rhizome shooting | Garg et al. 1967 |
| Low-T | *Agropyron repens* | in vivo | Whole plants | 10°C | Transplanting → 6th leaf | promotes rhizome initiation | McIntyre 1967 |
| Low-T | *Poa pratensis* | in vivo | Whole plants | 0–2 °C | 10–40 days | inhibits rhizome initiation | Moser et al. 1968 |
| Low-T | *Poa pratensis* | in vivo | whole plants | 0–2°C | 10–40 days | inhibits rhizome elongation | Moser et al. 1968 |
| Low-T | *Alstroemeria* | in vivo | whole plants | 9–13°C | 6–10 weeks | promotes rhizome branching | Vonk Noordegraaf 1981 |
| High-T | *Poa pratensis* | in vivo | whole plants | high day temperature (21 °C) | 9 weeks | promotes rhizome elongation | Aamlid 1992 |
| High-T | *Oryza longistaminata* | in vivo | whole plants | 28–30 °C | 3–4 weeks | promotes rhizome elongation | Wang et al. 2024a |
| High-T | *Nelumbo nucifera* | in vivo | whole plants | 20°C–30°C | 2 months | promotes rhizome branching | Masuda et al. 2006 |

| Stress | Species | Method | Material | Condition | Duration | Effect | Reference |
|---|---|---|---|---|---|---|---|
| High-T | *Cyperus esculentus* | in vivo | whole plants | 33/27 °C (day/night) | 28 days | promotes rhizome shooting | Garg et al. 1967 |
| Drought | *Leymus chinensis* | in vivo | whole plants | 30%–35% field capacity | 90 days | inhibits rhizome initiation | Wang et al. 2019 |
| Drought | *Carex lasiocarpa* | in vivo | whole plants | Soil water content (49%, w/w) | 40 days | inhibits rhizome initiation | Yuan et al. 2017 |
| Drought | *Chrysanthemum morifolium* | in vivo | whole plants | relative water content (50%–10%) | 30 days | inhibits rhizome initiation | Zhang et al. 2022 |
| Drought | *Triglochin buchenaui* | in vivo | whole plants | 200 mL/2 weeks, bottom–watering | 3 months | promotes rhizome initiation | Tabot and Adams 2012 |
| Drought | *Leymus secalinus* | in vivo | whole plants | −50% growing-season precipitation | 6 years | promotes rhizome initiation | Zheng et al. 2021 |
| Drought | *Leymus chinensis* | in vivo | whole plants | 30%–35% field capacity | 90 days | inhibits rhizome elongation | Wang et al. 2019 |
| Drought | *Carex lasiocarpa* | in vivo | whole plants | Soil water content (49%, w/w) | 40 days | inhibits rhizome elongation | Yuan et al. 2017 |
| Drought | *Chrysanthemum morifolium* | in vivo | whole plants | relative water content (50%-10%) | 30 days | inhibits rhizome elongation | Zhang et al. 2022 |
| Drought | *Leymus secalinus* | in vivo | whole plants | −50% growing-season precipitation | 6 years | promotes rhizome elongation | Zheng et al. 2021 |
| Drought | *Kohleria eriantha* | in vitro | rhizome | 1 mL water (low moisture); PEG −6 MPa | 20 days | inhibits rhizome initiation | Almeida et al. 2005 |
| Drought | rhizoma perennial peanut | in vitro | rhizome | 159 mm (Mar–May) | 80 days | inhibits rhizome shooting | Rice et al. 1996 |
| Drought | *Phragmites australis* | in vivo | whole plants | 35–40% field capacity | 120 days | inhibits rhizome shooting | Mingyang et al. 2022 |

| Drought | *Phragmites australis* | in vivo | whole plants | 35–40% field capacity | 90 days | promotes rhizome shooting | Mingyang et al. 2022 |
| Drought | *Festuca arundinacea* | in vivo | whole plants | Soil volumetric water content: 6.3% | 7 days | inhibits rhizome initiation | Ma et al. 2020 |
| Drought | *Festuca arundinacea* | in vivo | whole plants | Soil volumetric water content: 6.3% | 7 days | inhibits rhizome elongation | Ma et al. 2020 |